\documentclass[aps,showpacs,amsfonts,latexsym,amsthm,twocolumn]{revtex4}
\usepackage{amssymb}
\usepackage{graphicx}

\begin{document}

\newcommand{\be}{\begin{eqnarray}}
\newcommand{\ee}{\end{eqnarray}}
\newcommand{\bea}{\begin{eqnarray}}
\newcommand{\eea}{\end{eqnarray}}
\newcommand{\beq}{\begin{equation}}
\newcommand{\eeq}{\end{equation}}
\newcommand{\bma}{\begin{subequations}}
\newcommand{\ema}{\end{subequations}}
\newtheorem{lemma}{Protocol}
\def\bra #1{\langle #1\vert}
\def\ket #1{\vert #1\rangle}
\def\braket #1#2{\langle #1 \vert #2\rangle}
\def\ketbra #1#2{\vert #1\rangle \! \langle #2\vert}

\def\lR{l^2_{\mathbb{R}}}
\def\RR{\mathbb{R}}
\def\E{\mathbf e}
\def\D{\boldsymbol \delta}
\def\S{{\cal S}}
\def\T{{\cal T}}
\def\dd{\delta}
\def\one{{\bf 1}}
\def\eps{\varepsilon}

\title{Reversible universal quantum computation within translation invariant systems}

\author{K. G. H. Vollbrecht$^1$ and J. I. Cirac$^1$}
\affiliation{$1$ Max-Planck Institut f\"ur Quantenoptik,
Hans-Kopfermann-Str. 1, Garching, D-85748, Germany }

\pacs{03.75.Lm, 03.67.Lx, 42.50.-p}
\date{\today}

\begin{abstract}
We show how to perform reversible universal quantum computation on a translationally invariant pure state, using only global operations based on next-neighbor interactions. We do not need not
to break the translational symmetry of the state  at any time during the computation. Since the  proposed scheme fulfills the locality condition of a
quantum cellular automata, we present a reversible quantum cellular automaton capable of universal quantum computation.
\end{abstract}

\maketitle

\section{Introduction}

The great challenge in quantum information theory and quantum computation is to build a quantum computer, which is conjectured to
be exponentially faster than its classical counterpart. Many proposals have been presented in recent years that claim a physical
system to be a good candidate for doing quantum computation. In order to be able to build a quantum computer, the
proposed scheme has to satisfy a checklist \cite{DPV} of required properties that are widely
believed to be necessary for universal quantum computation. Two of the points  on this checklist are, that  a qubit has
to be encoded into a well defined physical system, i.e., there should be two levels
representing  $\ket{0}$  and  $\ket{1}$, and that each of these systems should be
manipulated individually. We present a quantum computation scheme that seems to get around these requirements, using translationally invariant states and
global addressing.

To encode several qubits into a translationally invariant system seems to be  contradictory, because by definition all individual systems
are identical and can not carry different kinds of quantum information. Restricting to  global operations that are themselves translationally invariant, we are not able to
break the translation symmetry, in particular, we can not address a single system alone.

But such a situation occurs for example in optical lattices, for which it is experimentally a very hard task to address single sites in the lattice.
In the ideal situation, this lattice is prepared in a translationally invariant state, with exactly one atom per site  \cite{bosonsol4}, all in the same internal state.
Schemes for quantum computation that have been proposed for this system require  breaking the translational symmetry as a first step \cite{KP, CDJ, IMN}  , e.g.,
 by  using imperfections in the lattice \cite{EQC}.

In this paper we introduce a novel method of performing quantum
computations on a translationally invariant, one dimensional lattice of 5-level systems, that can
be associated for example with a line of atoms in an optical lattice.
Our scheme will be completely based on global operations, that can be carried out by reflectionally symmetric and translationally invariant next-neighbor Hamiltonians.
Beside its physical relevance  these kind of operations allow us to connect  quantum computation
 to the context of quantum cellular automata \cite{qca, SL, SB}.

Our system will stay in a pure, reflectionally symmetric and translationally invariant state during the whole computation.
The main idea is based on the notion of ensemble quantum computation and is related to the schemes presented in \cite{EQC}.

The paper is organized as follows: in the first step we will introduce a quantum computation scheme that requires   non-translationally invariant states.
In the next step, we will show how we can get rid of this condition by switching to an ensemble quantum computation scheme. In a last step we will verify, that
we can do this with a pure state.

\section{Preliminaries and Notation}
 Assume we have atoms  with $6$ internal states
($\ket{0} \dots \ket{5}$) in a one dimensional lattice of length
$m$, where we assume $m$ to be very large. To avoid any discussion about border effects, we will further assume
either periodic boundary conditions or $m\rightarrow
\infty$.

$H_{xa}^{yb}$ denotes a reflection symmetric next-neighbor Hamiltonian, which is constructed in the following way:
$h_{xa}^{yb}=\ketbra{x,a}{y,b}+h.c.$ is a Hamiltonian having
the ability to transform $\ket{xa}$ into $\ket{yb}$ and vice versa, in the sense
that $e^{ih_{xa}^{yb}t} \ket{xa}=i\ket{yb}$ for  $t=\pi/2$. Here $x,y$ denote two states
in one site of the lattice and $a,b$ two states in a neighboring site. In most of the cases
$x,y,a,b$ will be chosen to be one of the basis states ($\ket{0} \dots \ket{5}$).

To make this  Hamilton reflection symmetric we take $h_{(xa,yb)}=h_{xa}^{yb}+h_{ax}^{by}$.
To our system we will apply  translation invariant next neighbor Hamiltonians of the form
\beq \label{H}
H_{xa}^{yb}=\sum_{i} h_{(xa,yb)}^{(i,i+1)} ,\eeq
where $h_{(xa,yb)}^{(i,i+1)}$
 denote the
Hamiltonian $h_{(xa,yb)}$ applied on the $i$th and $(i+1)$th place.
In particular, we will only apply Hamiltonians of the form $H_{xa}^{xb}$, with $\braket{x}{a}=\braket{x}{b}=0$.
In this case all $h_{(xa,xb)}^{(i,i+1)}$ commute with each other. Note, that the state $a,b$ do not have to be orthogonal.

By $U_{xa}^{xb}$ we denote the unitary operation $e^{i H_{xa}^{xb} t}$ with
$t=\pi/2$
 is chosen in such a way that every  "isolated" $\ket{xa}$ in the lattice is transformed into  $\ket{xb}$ and vice versa, in the sense that
$U_{xa}^{xb} \ket{\dots xac\dots }=i \ket{\dots xbc \dots }$ ($c \neq x$).
Without loss of generality we will ignore the factor i in the following.
Since all terms in (\ref{H}) commute (for $x=y, x \neq a,b$) we can calculate the action of $U_{xa}^{xb}$ onto the lattice
by applying twice the unitary operation
\beq U_{xab}\otimes U_{xab} \dots \otimes U_{xab}\eeq
but the second time shifted by one lattice site, where
$U_{xab}$ is a two qubit unitary operation defined as
$$ U_{xab}\ket{xa}=\ket{xb}, U_{xab}\ket{xb}=\ket{xa},$$
$$U_{xab}\ket{ax}=\ket{bx}, U_{xab}\ket{bx}=\ket{ax}. $$
Note, that this implies a finite propagation speed of quantum information, of one lattice site per operation, within the lattice, i.e.,
any manipulation on one site can effect another site after $x$ operations at the earliest, if the two sites are $x$ sites apart. Note that this is exactly
the kind of condition that defines  a system to be a quantum cellular automaton \cite{qca}.
By $V_{xy}$ we denote a unitary operation, that exchanges the levels $x$ and $y$ in every site of the lattice, i.e., $V_{xy}=\tilde{V}_{xy}^{\otimes m}$, with
$\tilde{V}_{xy}\ket{x}=\ket{y}, \tilde{V}_{xy}\ket{y}=\ket{x}$.

\section{a quantum Computer scheme}
Now we show how to carry out a quantum computation using only translation and reflection invariant next-neighbor interactions, e.g.,
using only operations of the form $U_{xa}^{xb}, V_{xy}$,
but
assuming, in a first step, that the lattice is prepared in the non-translation invariant state
\beq
\ket{23000000000000000 \dots}.
\eeq
The $\ket{23}$ at the beginning is called the pointer, every following "zero"  will considered to be a qubit with the possible values $\ket{0},\ket{1}$.
The trick of doing quantum computation in this picture is that we will use Hamiltonians that somehow are located  due to the pointers, e.g. the unitary
$U_{30}^{31}$ have only an effect to the qubit next to the "$3$" pointer. For the above state  only the first  qubit is affected, the state transforms into
\beq
\ket{231000000000000000 \dots}.
\eeq

To address other qubits than the first one, we have to "move" the pointer around. This can be done in the following way:
\begin{itemize}
\item
moving the pointer one site  to the right (one extra level $4$ required):
By the sequence $S_\rightarrow$:
$U_{03}^{04}$,
$U_{40}^{42}$,
$U_{04}^{03}$,
$U_{13}^{14}$,
$U_{41}^{42}$,
$U_{14}^{13}$,
$V_{23}$,
the pointer $\ket{23}$ is shifted one position to the right.
The qubit right of the pointer is shifted at the same time two sites to the left.

To shift the pointer one position to the left we just apply the protocol backwards.

\end{itemize}
By moving the pointer to a special qubit and using Hamiltonians that are "located" by the pointer, we can apply arbitrary unitary operations to any qubit.
\begin{itemize}
\item local operations: First we  move the  pointer to the position left of the  qubit we want to adress.
 Then we apply Hamiltonians of the form $H_{3x}^{3y}$, where $x$,$y$ can be any qubit state $\alpha\ket{0}+\beta\ket{1}$. Doing this we can
 apply any one-qubit-unitary on the qubit .
\end{itemize}
To do quantum computation we need controlled operations, like a C-NOT gate. For this task, it is sufficient to have a C-NOT gate acting only between neighboring sites.
\begin{itemize}
\item controlled operations (CNOT) between two neighboring sites:
By the sequence
$U_{02}^{04}$,$U_{43}^{42}$,$U_{21}^{20}$,$U_{43}^{42}$,$U_{21}^{20}$,$U_{02}^{04}$
we apply a CNOT gate between the two qubits on the left and on the right site of the pointer.
The qubit lying near the "2" is the source, the qubit near the "3" the target. To exchange the role of source
and target we just have to exchange the role of "2,3" in the sequence. Since the pointer can be moved, we can apply this
operation to arbitrary neighboring two-qubits.
\end{itemize}
Local operations and CNOT between two neighboring sites form a universal set of quantum gates, which enable us to do any kind of quantum
computation.
Finally, we need to measure a single qubit.
\begin{itemize}
\item measurement: First we move the pointer to the qubit we want to measure. Then we apply $U_{31}^{34}$ and count the sites found in $\ket{4}$
, i.e., we will find either zero ore one atom in state $\ket{4}$.
\end{itemize}

\subsection{ensemble quantum computation}
Using this kind of quantum computation scheme we can do ensemble quantum computations, i.e., we can run several quantum computers in parallel.
If we start with several pointers, e.g. the state
$$\ket{23000000000230000000023000000 \dots}$$
we will have three quantum computers running in parallel. We only have to take care, that the distance between two pointers is larger than
the number of qubits $n$ used in (one copy) of the quantum computer.

In an ensemble quantum computation scheme a measurement takes place in all quantum computers at the same time.
The number of atoms found in state $\ket{4}$ is given by the expectation value
$<M_4>=tr \rho M_4 $  and can be compared to the number of atoms found in state $\ket{3}$, $<M_3>=tr \rho M_3 $ , i.e.,
to the number of quantum computers, with
\beq
M_x=\sum_{i} P_x^{(i)},
\eeq
where $P_x^{i}=\ketbra{x}{x}$ applied on the $i$th site.
Note that this measurement is again a global
operation.

We can use pointers of the form $\ket{23}$ and $\ket{32}$ at the same time, e.g.
$$\ket{2300000000000000000003200000000023000000 \dots}.$$
The pointers of the form $\ket{23}$ will address the qubits to their right, the  $\ket{32}$ the qubits
to their left.

\subsection{Starting from a mixed translational invariant states}
We will now show, that we can perform  ensemble quantum computation starting from a translation invariant state.
For simplicity, we first start with a mixed state, where every site
is either in $\ket{0}$ or in $\ket{5}$, with corresponding probabilities
$1-\eps$, $\eps$. So we start with a mixed state
\beq \label{mix}
\rho=\sum_{i} p_i \ketbra{\phi_i}{\phi_i},
\eeq
where $\ket{\phi_i}$ is the set of all possible states only consisting of $\ket{0}$ and $\ket{5}$, e.g.,
the state
\beq
\ket{0000005000000000000050005000000005} . \label{example}
\eeq
The sum goes over all possible configurations of $\ket{5}$ and $\ket{0}$, with the corresponding
probabilities $p_i=\eps^{\#\ket{5}}(1-\eps)^{\#\ket{0}}$.
This mixed states is a translation invariant and can be for example produced from the state $\ket{0}^{\otimes n}$ by
adding some incoherent translation invariant noise that is mixing $\ket{0}$ and $\ket{5}$.

To do quantum computation using above scheme
we need to create pointers.

\subsection{Creating pointers}

By the sequence $U_{50}^{52}$,$U_{20}^{23}$,$U_{32}^{34}$, $U_{50}^{52}$ and $U_{32}^{34}$, we create $\ket{23}$ and $\ket{32}$ pointers.
The $\ket{5}$ states cuts the lattice in partitions of different length. Every such partition with more than 4 $\ket{0}$ gets a $\ket{23}$ and $\ket{32}$ at the  borders,
while partitions  with less than 3 $\ket{0}$ stays unchanged. E.g., state (\ref{example}) transforms into
\beq
\ket{0000325230000000003250005230000325}.
\eeq
Each large enough partition contains two quantum computers; each of them can address half of the qubits in the partition. Note, if you move the pointers too far, so that
they meet each other, they just cross each other. If the length of the partition  is even and we apply $S_\rightarrow$ on $\ket{2332}$ we get
$\ket{3223}$. In the odd case, the meeting pointers first get into an "inactive" state, and in the next step pass each other. In particular, if we apply
two times $S_\rightarrow$, we get
$\ket{23032}\rightarrow\ket{04040}\rightarrow\ket{32023}$ and
$\ket{23132}\rightarrow\ket{14141}\rightarrow\ket{32123}$.
Note, that the states $\ket{04040},\ket{14141}$ are "inactive", in the sense that they are not affected if we apply a CNOT or a local operations with the
Hamilton $H_{3x}^{3y}$.

 If we try to move the pointer over the $\ket{5}$ border the pointers change their direction, i.e., $\ket{235}\rightarrow\ket{325}$. During any kind of quantum computation,
 the pointers can not leave their partition and the total number of pointers stays constant.

\subsection{Signal and scalability}
Every $\ket{00500}$ splits up into two quantum computers "23" and "32", one running to the left and one running to the right.
If we use this for ensemble quantum computation with a algorithm using $n$ qubits, we get in the end of the protocol a correct
signal from all partitions of length greater than $2n+4$. On smaller partitions the algorithm will not run correct, because the two pointers
in the partition will address qubits belonging to the other pointer.

 We can control the number
of correctly working quantum computers and incorrectly working quantum computers by the probability $\eps$.

The expected total number of partitions found in a lattice of size $m$ generated in this way is given by
$m \eps$. Each of this partitions has a random length, where
the probability having exactly $n$ $\ket{0}$-sites is given by $(1-\eps)^{n}\eps$. The probability that a partitions
has $n$ or more sites is given by  $(1-\eps)^{n}$.

Therewith, the  expected number of  partitions of length bigger
than $2n+4$ on a $m$ site lattice is given by
$$\#working=m \eps(1-\eps)^{2n+4},$$
where $m \eps$ is the total number of partitions found in the lattice  and $(1-\eps)^{2n+4}$ is the probability that  such a  partition is
bigger than $2n+4$. The ratio of long enough partitions to  total number
of partitions is then just $(1-\eps)^{2n+4}$.
If we choose the probability $\eps$ small enough, then the ratio  is going to one, i.e.,
the signal of the measurement will only come from correctly working quantum computers.
But choosing $\eps$ to be very small can result in a scalability problem for our scheme, because $\eps$ is  responsible
for the total number of quantum computers found in the lattice. Fortunately, we have to decrease $\eps$ only polynomially with the number qubits $n$.
Choosing $\eps\sim 1/n^2$ we get
$$\#working = m\frac{1}{n^2}(1-\frac{1}{n^2})^{2n+4}\sim m\frac{1}{n^2}.$$
This means, that the
probability of a quantum computer to be working goes to one, while the total number of quantum computers decrease only polynomially with $\frac{1}{n^2}$.

\subsection{Starting from pure translational invariant states}
Instead of using a mixed translation invariant state, we can start with the pure state
\beq
\ket{0}^{\otimes m}.
\eeq
This can be transformed into
\beq
\ket{\Phi}=(\sqrt{1-\eps}\ket{0}+\sqrt{\eps}\ket{5})^{\otimes m}.
\eeq
by applying the same unitary on every site, which is obviously a translation invariant unitary operation.
We claim, that the above procedure starting with a translational invariant mixed state works in exactly the same way for this translation invariant pure state.

Instead of having a mixture of states, we now have the coherent superposition of exactly the same states, i.e.,
\beq \label{pure}
\ket{\Phi}=\sum_i \sqrt{p_i} \ket{\phi_i},
\eeq
where $p_i$ and $\ket{\phi_i}$ are as in (\ref{mix}). During a quantum computation the state transforms to
\beq
\ket{\Phi'}=\sum_i \sqrt{p_i} \ket{\phi'_i}.
\eeq
For the final measurement we get
\beq
\braket{\Phi'}{M_x \Phi'}=\sum_{ij} \sqrt{p_i p_j} \braket{\phi'_i}{M_x \phi'_j}=\sum_{i} p_i \braket{\phi'_i}{M_x \phi'_i}.
\eeq
The second equality is due to the fact, that the operations and  measurements do not affect the $\ket{5}$ levels. This guarantees that $\braket{\phi'_i}{M_x \phi'_j}=\delta_{ij} \braket{\phi'_i}{M_x \phi'_i} $.
The measurement gives the same results, as if we would have used  the  state (\ref{mix}), for which we already proved our scheme to work.

\subsection{Using only 5 internal levels.}
We can reduce our scheme using only $5$ internal states. For preparing the pointer
we just replace the role of the $\ket{5}$ by a $\ket{1}$, i.e., we transform randomly distributed $\ket{00100}$ into $\ket{32123}$.  Using this kind of scheme the pointers
can break out of their partition and effect the computation in neighboring partitions. But this turns out to be no problem, because by  choosing $\eps\sim 1/n^2$ the probability
for this goes to zero compared to the probability of correctly working quantum computers.

Using the pure state scheme we get the problem, that we  can not ensure any more, that $\braket{\phi'_i}{M_x \phi'_j}=\delta_{ij}\braket{\phi'_i}{M_x \phi'_i}$, since the $\ket{1}$ configuration, that
defines the states, can be changed due to the incorrect working quantum computers. Instead of using the $\ket{1}$ configuration, we take now the configuration
of the pointers and use the same kind of argumentation.
Let us take the state
\beq
\ket{\Phi}=(\sqrt{1-\eps}\ket{0}+\sqrt{\eps}\ket{1})^{\otimes m}.
\eeq
and apply the pointer creation protocol $U_{10}^{12}$,$U_{20}^{23}$,$U_{32}^{34}$, $U_{10}^{12}$ and $U_{32}^{34}$.
We can write the resulting states as
\beq \label{pure2}
\ket{\Phi}=\sum_i \sqrt{q_i} \ket{\psi_i},
\eeq
where $i$ labels all possible pointer configurations that can occur due to the protocol, i.e.,
all possible configurations of $\ket{23},\ket{32}$. This is basically the same
decomposition of the state $\ket{\Phi}$ as in (\ref{pure}), with the difference that all sets of states
$\{\ket{\phi_i}\}$ that differ only at
partitions that are too small to create any  pointer are merged together to one $\ket{\psi_k}\sim \sum_i p_i \ket{\phi_i}$,
e.g. the two states
$$\alpha \ket{\dots00321112300\dots}, \beta \ket{\dots00321012300\dots}$$
have different $\ket{1}$ configuration, i.e., they correspond ($\ket{1}$=$\ket{6}$) in (\ref{pure}) to to differen $\ket{\phi_i}$.
In (\ref{pure2}) they are represented now in only one term,

$$\ket{\dots0032}\ket{\alpha 101+ \beta 111}\ket{2300\dots},$$
because they have the same pointer configuration. All working quantum computers
in such a $\ket{\psi_i}$ are correctly initialized with only zeros.
Superpositions, like $\ket{\alpha 101+ \beta 111}$, appear only in
areas that are not addressed by a working quantum computer.
There are no pointers destroyed during the computation
or the measurement. This guarantees  now,
that we get $\braket{\psi'_i}{M_x \psi'_j}=\delta_{ij}\braket{\psi'_i}{M_x \psi'_i}$ and therewith
\beq
\braket{\Phi'}{M_x \Phi'}=\sum_{i} q_i \braket{\psi'_i}{M_x \psi'_i}.
\eeq
The measurement outcomes are the same as if we would have started with the mixed state
$\rho=\sum_i q_i \ketbra{\psi_i}{\psi_i}$. The number of correctly working to incorrectly working quantum computers is the
same as in (\ref{mix}), i.e., we can effectively suppress the signal coming from incorrectly working quantum computers.

\section{Conclusion}
We have presented a scheme to perform universal quantum computation within a translationally and reflection invariant system.
This scheme requires 5 internal levels at each position, global-local operations and next-neighbor interactions.
The presented scheme is scalable in the sense that the number of resources, i.e. sites, scales quadratically
 \footnote{Note, that so far we assume  no errors occurring during the quantum computation.}
 with the number of qubits used during the quantum computation.
We formulate our scheme in the language of Hamiltonians, but we could have started from unitaries which
are translationally and reflection invariant.
Apart from the obvious relation to some physical implementation, like atoms in optical lattices, our result shows it
is possible to perform quantum computation in translationally invariant systems.
Furthermore, in the case of unitaries, what we have built is a quantum cellular automaton.
Thus our result implies that the power of a quantum cellular automata is equivalent to that
of quantum computers.
An open question is whether it is possible to achieve the same results
with qubits, or if it is strictly necessary to use more than two levels per
site. In the latter case, this would show that higher dimensional systems are indeed more
powerful than qubits for certain quantum informational tasks.

We thank  E. Solano for useful discussion. This work was supported by EU through RESQ and QUPRODIS projects and
the Kompetenznetzwerk Quanteninformationsverarbeitung der Bayrischen Staatsregierung.

\end{document}